\documentclass[bibnotes,twocolumn,aps,prl,a4paper]{revtex4-1}

\usepackage{color,epsfig,rotating,graphicx,subfigure}
\usepackage{amsmath,amssymb,amscd}

\usepackage[latin1]{inputenc}
\usepackage[english]{babel}

\usepackage{dsfont}
\usepackage{textcomp}

\begin{document}

%
%

\title{Diamond particles as nanoantennas for nitrogen-vacancy color centers}

\author{J.-J. Greffet$^1$}
\author{J.-P. Hugonin$^1$}
\author{M. Besbes$^1$}
\author{N. D. Lai $^2$}
\author{F. Treussart$^2$}
\author{ J.-F. Roch$^2$}
\affiliation{\\
\\$^1$
Laboratoire Charles Fabry, 
Institut d'Optique, Univ Paris Sud, CNRS UMR 8520,
2 av Fresnel, 91127 Palaiseau Cedex, France
\\$^2$Laboratoire de Photonique Quantique et Mol\'eculaire \& CNRS -- UMR 8537,\\
\'Ecole Normale Sup\'erieure de Cachan,  61 avenue du Pr\'esident Wilson, F-94230 Cachan cedex, France
}

\begin{abstract}
The photoluminescence of nitrogen-vacancy (NV) centers in diamond nanoparticles exhibits  specific properties as compared to NV centers in bulk diamond. For instance large fluctuations of  lifetime and brightness from particle to particle have been reported. It has also been observed that for  nanocrystals much smaller than the mean luminescence wavelength, the particle 
 size sets a lower  threshold for resolution in Stimulated Emission Depletion (STED) microscopy. We show that all these features can be quantitatively understood by realizing that the absorption-emission of light by the NV center is mediated by the diamond nanoparticle which behaves as a dielectric nanoantenna. 
\end{abstract}

\maketitle

\newpage

%
%

The  negatively charged nitrogen-vacancy (NV$^{-}$) color center in diamond is a solid-state artificial atom with unique room-temperature properties. Its perfectly stable photoluminescence and an optically detectable electron spin lead to a wide range of applications such as efficient on-demand generation of single photons \cite{Beveratos2002} and highly sensitive nanoscale magnetic and electric sensing \cite{Maze2008,Balasubramanian2008,Dolde2011}. These properties are preserved when NV centers are created in diamond particles \cite{Beveratos2001,Treussart2006} with size down to the nanometer range \cite{Tisler2009,Rondin2010}. The photoluminescence of a single NV center in a nanocrystal was used for generating single plasmons once coupled to metallic structures \cite{Kolesov2009,Cuche2010}. Nanodiamonds embedding many NV centers have been developed for bioimaging purposes \cite{Chang08,Faklaris2009}. Moreover, due to its perfect photostability, the NV color center can be revealed using far-field nanoscopy, based on either stimulated emission depletion (STED) or ground-state depletion (GSD) microscopy techniques \cite{Rittweger2009a, Rittweger2009b, Han2009, Yan}. A 6-nanometer record resolution   has been achieved in the case of a NV defect in a bulk diamond sample \cite{Rittweger2009a}. These results promise applications of NV-based luminescent nanodiamonds for bio-imaging, with the potential of challenging the use of fluorescent molecules which suffer from limited photostability. 

All these applications require to understand how the optical properties of NV color centers are modified once nestled in a nanodiamond. Specific nanoparticle effects have been observed. For instance, the NV luminescence decay time exhibits a broad statistical distribution when recorded from a set of nanodiamonds spincoated on a coverglass and the lifetime is on average longer than the value measured for a NV center in a bulk diamond \cite{Tisler2009}. This lengthening  has been qualitatively attributed to modifications of the dielectric environment  \cite{Beveratos2001,Tisler2009}. It has also been noted that the lifetime distribution is not correlated to the brightness of the emitter \cite{Tisler2009,Han2009}. Finally, although a 6 nm resolution was reached in bulk diamond with STED imaging, the image of a single NV center in a $40$ nm nanodiamond was limited  by the size of the hosting particle \cite{Han2009,Yan}. 
 
In this letter, we show that the differences between emission by a NV center in a bulk or in a nanoparticle can be understood by considering that the dielectric nanoparticle acts as a dielectric nanoantenna similarly to diamonds nanopillars \cite{Loncar}. Furthermore, we demonstrate that the optical resolution of far-field nanoscopy techniques like STED imaging  is intrinsically limited to the nanodiamond size when considering the regime of particle dimension smaller than the emission wavelength. 

\begin{figure}[Hhbt]
  \epsfig{file=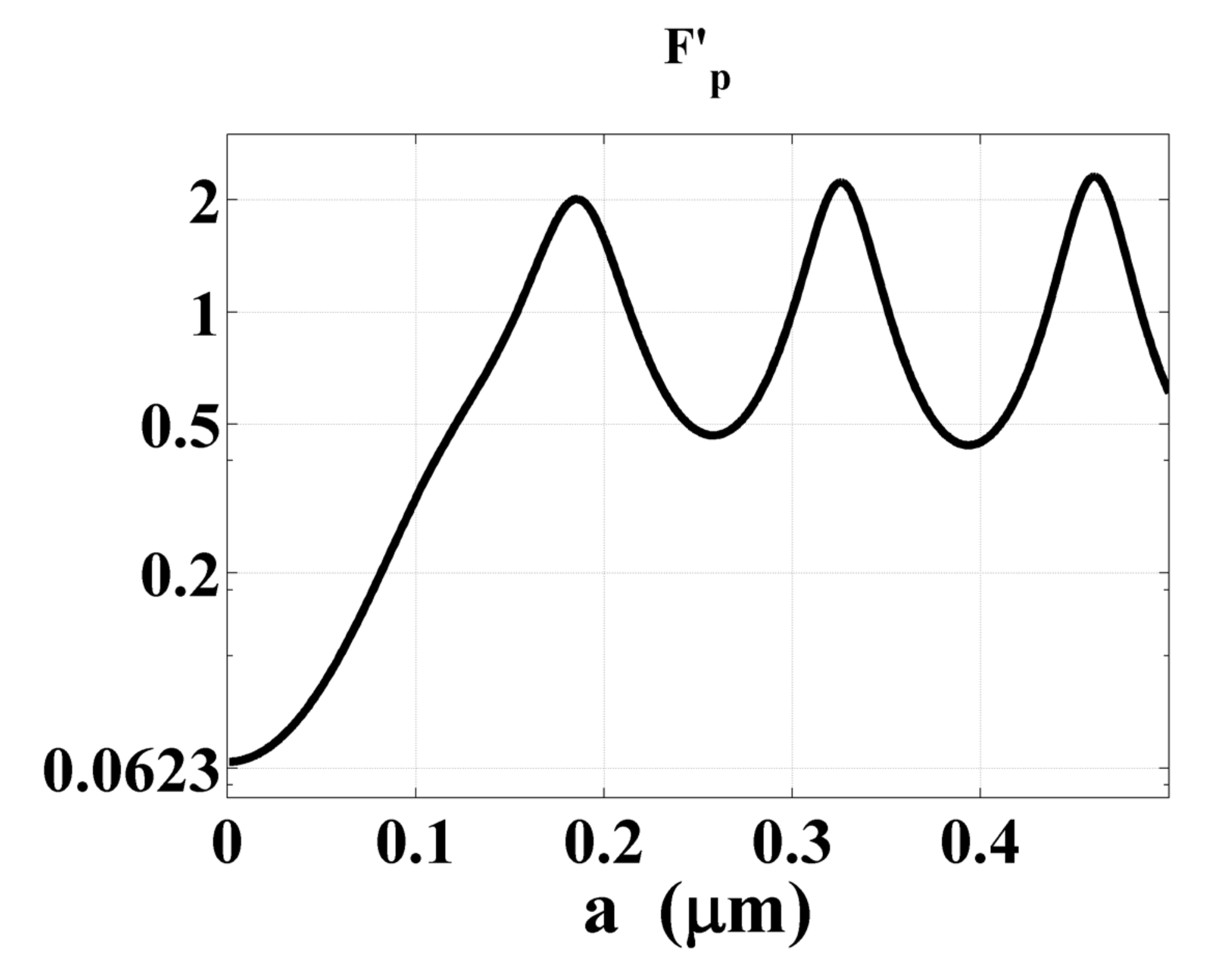, width=0.30\textwidth}
  \caption{Purcell factor associated with radiative emission of a NV center located at the center of a spherical nanodiamond for different radii a.}
  \label{FigPurcell}
\end{figure}

The optical properties of the NV center in the nanodiamond can be described using a classical electric dipole in a lossless dielectric particle of index of refraction $n$. To study the influence of the nanoparticle on its emission lifetime, we compute the ratio of  the total power emitted by a point-like monochromatic electric dipole in a nanosphere and in a bulk host. This ratio is equal to the emission rate $\Gamma_{\rm np}$ in a nanoparticle normalized by the emission rate $\Gamma_{\rm b}$ in the bulk. The calculations are done using the formalism developed by Mie \cite{Bohren,Druger,Gersten,Lange}.  A similar discussion has been reported for an ensemble of fluorophores embedded in a dielectric particle \cite{Gersten,Sandoghdar}. In such configurations, the question of the local field due to the near-field environment of the fluorophore plays a key role. Here, we consider a single NV center in a diamond nanosphere. We note that the local field correction is an irrelevant question as the NV center cannot be defined independently of the vacancy in the diamond lattice. Hence, we have normalized the emission rate of a NV center in a particle by the emission rate of the same NV center in a bulk diamond. In the absence of non-radiative decay, this ratio is equal to the ratio of local density of states (LDOS) \cite{Greffet}. Although a diamond nanoparticle is not a microcavity sustaining a single mode, we will refer to this normalized LDOS as Purcell factor $F^\prime_{\rm p}$, the prime indicating that emission in bulk diamond is taken as a reference. The Purcell factor is displayed in Fig.~\ref{FigPurcell} in the case of a dipole located at the center of the nanosphere when varying the radius $a$ of the nanoparticle. We observe that $F^\prime_{\rm p}<1$ for particles in the so-called electrostatic regime, i.e. $a\ll\lambda$ where $\lambda$ is the vacuum wavelength at the dipole emission frequency. When the particle radius becomes larger than $\lambda/n$, the Purcell factor first increases and then oscillates due to Mie resonances in the sphere \cite{Gersten,Sandoghdar}.

\begin{figure}[Hhbt]
  \epsfig{file=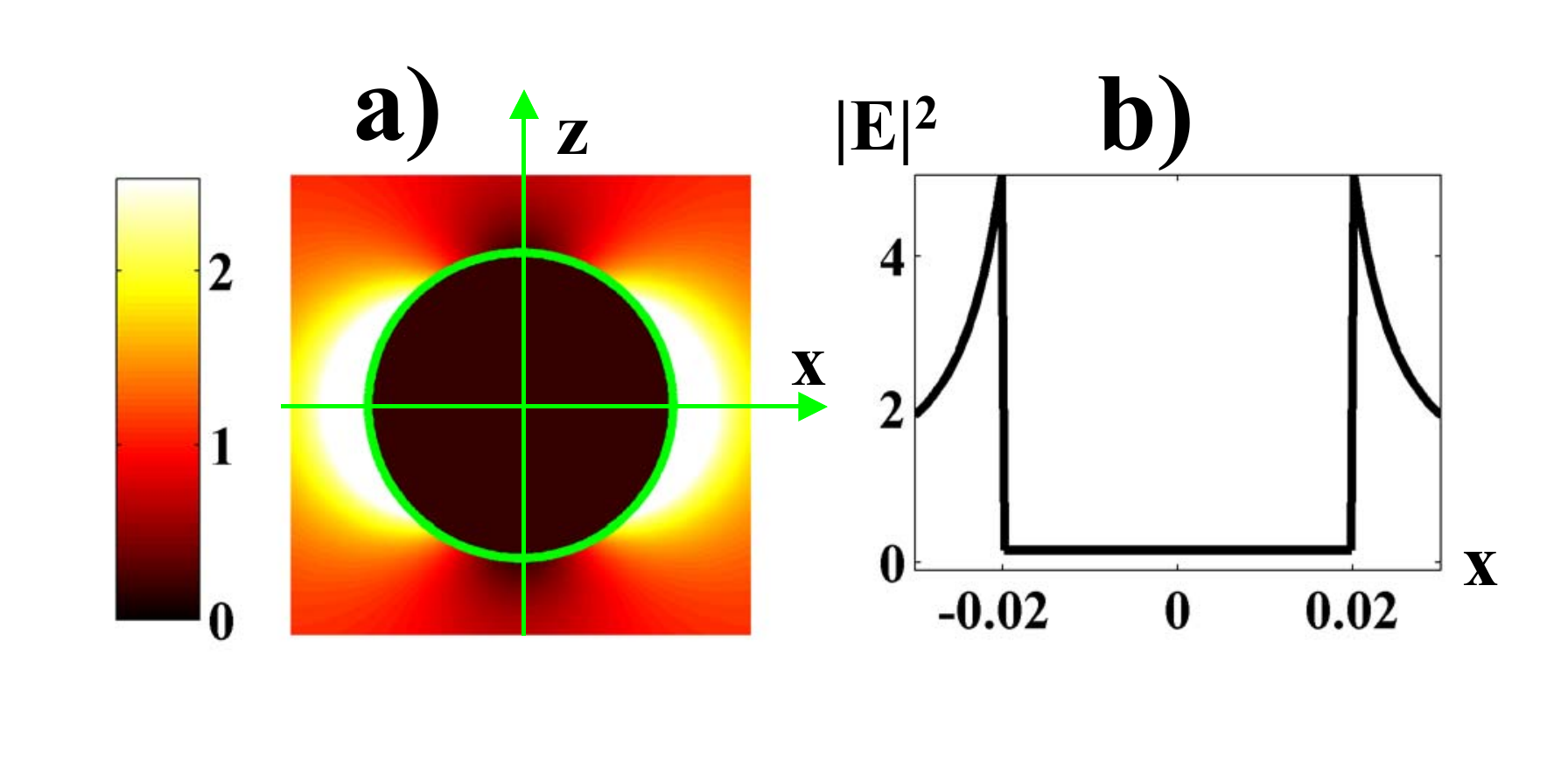, width=0.45\textwidth}
  \caption{Coupling between an embedded dipole in a 20 nm radius diamond nanoparticle and the far field. a) Square modulus of the electric field inside the sphere excited by a plane wave at 637 nm wavelength, corresponding to the zero phonon line of the NV$^{-}$ emission.  b)  Square modulus of the field along the $x$ axis.}
  \label{Intensity_nanosphere}
\end{figure}

Let us now analyse in more details the influence of the particle size on the emission rate. To understand the emission rate reduction in the electrostatic regime, we invoke the reciprocity theorem. In simple words, the reciprocity theorem states that a detected amplitude does not change upon exchange of the positions of a source and a detector \cite{Landau}. Hence, the far-field amplitude radiated in direction $\mathbf{u}$ by a dipole located at $\mathbf{r}$ inside the particle is proportional to the field created at  $\mathbf{r}$ by an incident plane wave illuminating the particle propagating along $-\mathbf{u}$. It follows that we can replace a far-field radiation problem by the computation of the field in the particle illuminated by a plane wave.  Fig. \ref{Intensity_nanosphere} shows the intensity $| \mathbf{E}|^2$ in the particle with $20$ nm radius illuminated by a plane wave  linearly polarized along Ox.

It is seen that the field is uniform within the sphere and much smaller than the incident field. This means that the Purcell factor does not depend on the position of the dipole in the particle. Furthermore,  for a dipole at the center, the spherical symmetry entails that the Purcell factor does not depend on polarization. Noting that for a small particle, the electrostatic approximation is valid, it follows that the internal field intensity is given by $\displaystyle \left\vert \frac{3}{n^2+2} E_{\rm inc} \right\vert^2$ \cite{Landau} where $E_{\rm inc}$ is the incident field amplitude. The low value of the field in a diamond nanoparticle is a manifestation of dielectric screening. To recover analytically the Purcell factor, we start by introducing the fictitious emission rate $\Gamma_{\rm vac}$ of a NV center in a vacuum as a non-physical but convenient intermediate parameter. The emission rate of a dipole in a bulk dielectric of index of refraction $n$ is related to $\Gamma_{\rm vac}$ by the relation $\Gamma_{\rm b}=n\Gamma_{\rm vac}$ which accounts for the ratio of densities of states.  Using the reciprocity argument, the emission rate in the nanosphere is related to the emission rate in a vacuum by accounting for the dielectric screening factor $\displaystyle \Gamma_{\rm np}=\Gamma_{\rm vac} \left(\frac{3}{n^2+2}\right)^2$.  It follows that $\displaystyle F^\prime_{\rm p}=\frac{1}{n}\left[\frac{3}{n^2+2}\right]^2$. Using $n=2.4$, we find $F^\prime_{\rm p}=0.0623 $ in excellent agreement with the numerical simulation shown in Fig.\ref{FigPurcell}. Finally, we note that a real NV center is described by two orthogonal emission dipole moments \cite{Ref_NV_Deuxdipoles}. As the lifetime does not depend on the dipole moment orientation in the  spherical case, this specific structure has no influence.

\begin{figure}[Hhbt]
  \epsfig{file=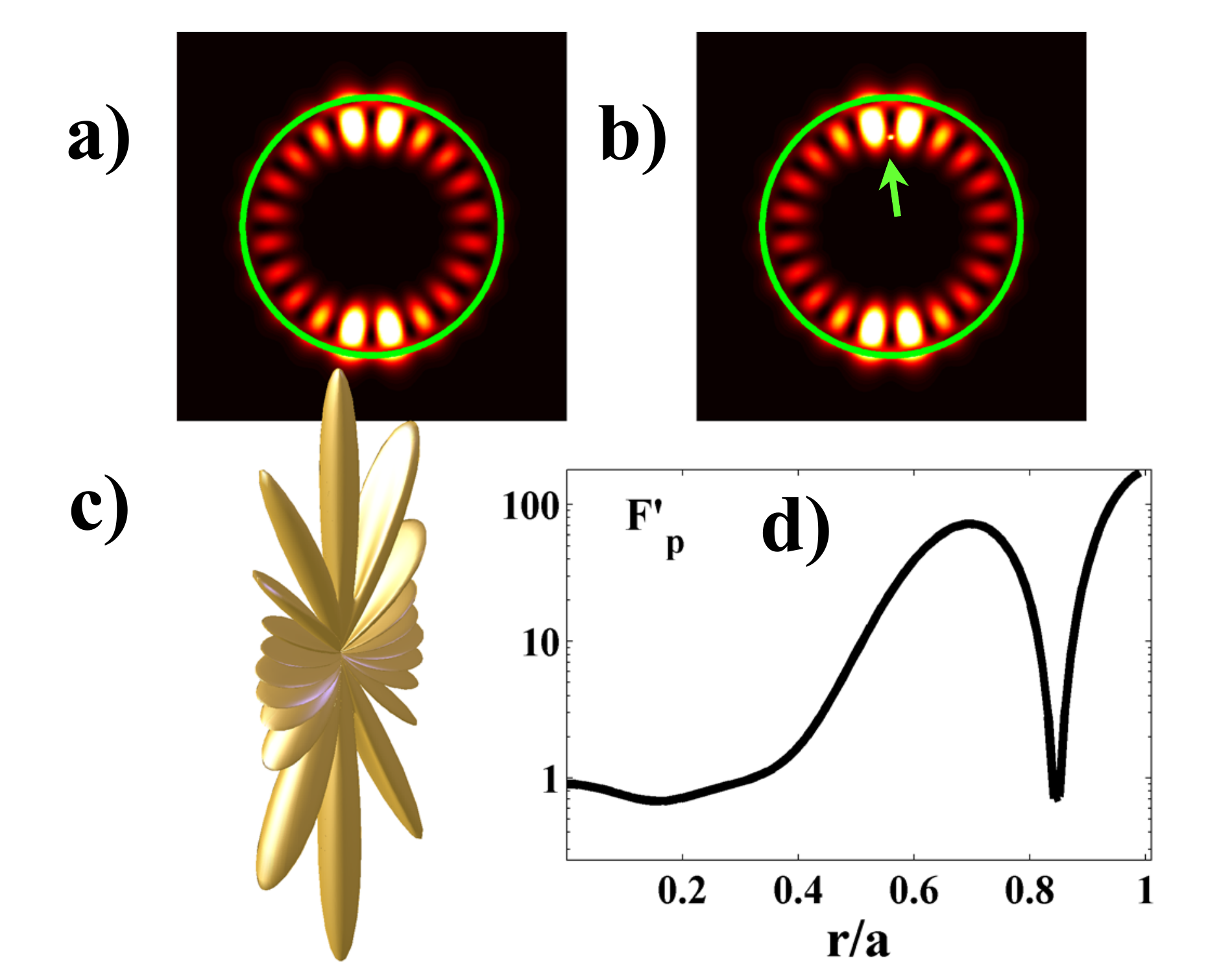, width=0.35\textwidth}
  \caption{Resonant excitation of a single mode structure in a  diamond nanoparticle of $ a=621.27$ nm radius. (a) Square modulus of the electric field in a diamond sphere excited by a plane wave. (b) Square modulus of the electric field excited by a dipole. The arrow indicates its position. (c)  angular emission pattern for an orthoradial dipole in the particle, (d) Purcell factor versus emitter position (with orthoradial polarization) .}
  \label{Fig_Mie_resonant}
\end{figure}

We now turn to the case of particles supporting Mie resonances. Fig. \ref{Fig_Mie_resonant}(a) shows the square modulus of the electric field within the particle when it is illuminated by a monochromatic plane wave tuned to a Mie resonance with a quality factor $2\,10^4$. As expected, the spatial structure of the field in the particle reproduces the mode field (not shown). According to the reciprocity theorem, if the source dipole is located at a maximum of the field, its radiation will be enhanced.  We plot in Fig. \ref{Fig_Mie_resonant}(b)  the field produced by the dipole when it is located at the point indicated by the arrow. Noting that the field structure matches the spatial pattern of the resonant mode field, we  conclude that the dipole radiation is mediated by the particle mode. Hence, emission of the NV center in the sphere is a two-step process: i) the dipole excites resonantly a mode, ii) the mode radiates in the vacuum. The far-field angular emission pattern is shown in Fig. \ref{Fig_Mie_resonant}(c) with an angular oscillation which again reproduces the angular pattern of the mode in the sphere. It follows that the brightness in a given solid angle can be almost null although the Purcell factor can be large. Hence, for a single emitter, one should not expect a direct correlation between the brightness in a given direction and the lifetime in the case of a particle large enough to support Mie resonances. Obviously, this is no longer correct if all the radiation is collected using e.g. an integrating sphere or for an ensemble of emitters in a sphere. 

From the reciprocity argument, we know that the strength of the coupling of the emitting dipole to the mode depends on the amplitude of the mode field at the dipole source position. Hence, we expect a strong dependence of the Purcell factor on the position of the NV center in the particle, as shown in Fig. \ref{Fig_Mie_resonant}(d).  As expected from Fig. \ref{Fig_Mie_resonant}(a), the Purcell factor is weak when the emitter is in the center of the particle and becomes large when the emitter is located close to a maximum of the mode. In summary, our analysis in the Mie regime shows that the emission of the NV center is mediated by the modes of the diamond particle. This nanoantenna effect can efficiently control the lifetime and the angular emission of the point defect in the nanodiamond.

\begin{figure}[Hhbt]
   \epsfig{file=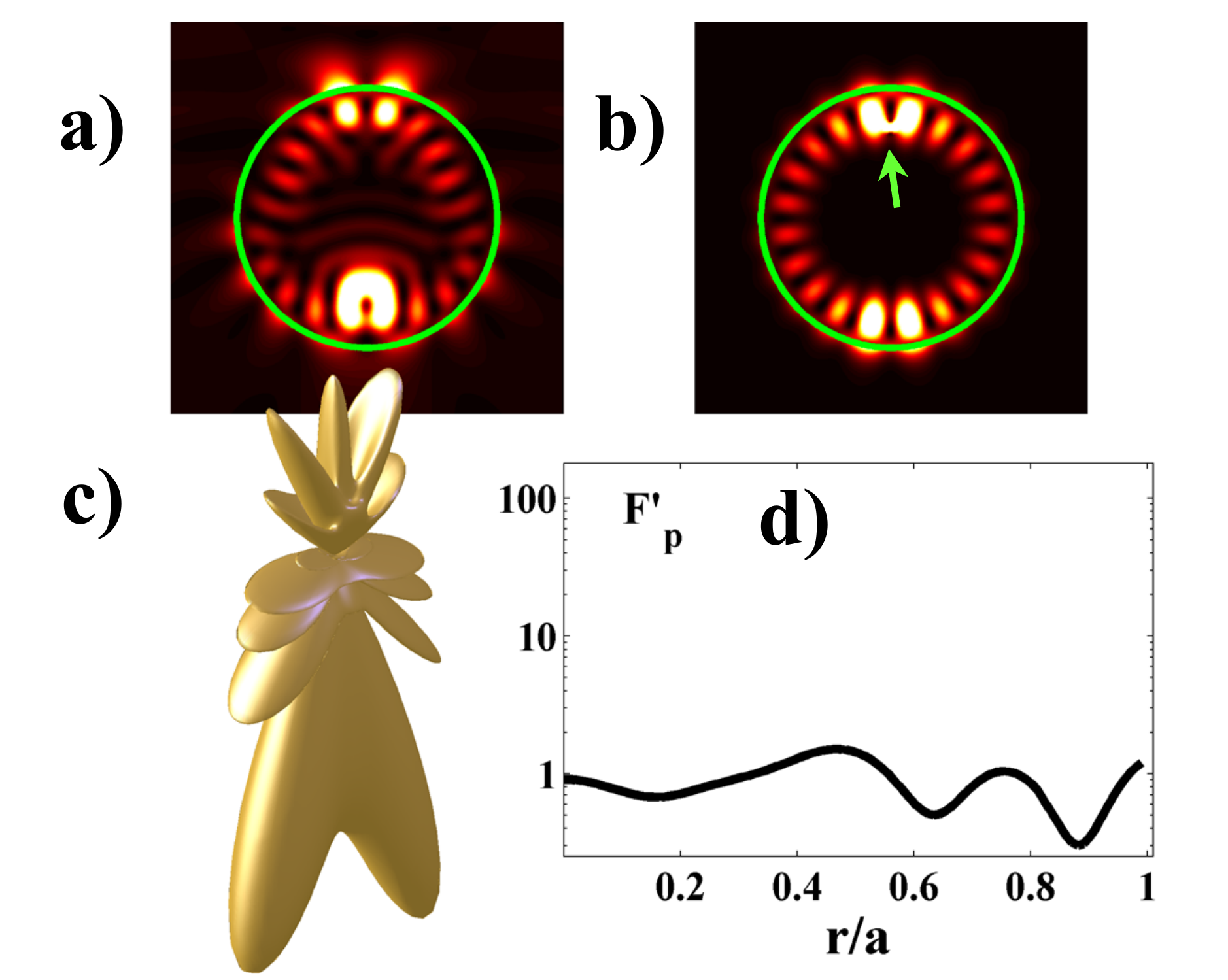, width=0.35\textwidth}
  \caption{Same as Fig. \ref{Fig_Mie_resonant}  but with $ a=621$ nm.  The slight detuning  drastically modifies the excitation of the modes (a,b) and consequently the angular emission pattern (c) and the Purcell factor (d).}
    \label{Fig_Mie_detuned}  
\end{figure}

We emphasize that Fig. \ref{Fig_Mie_resonant}  corresponds to the specific case for which the dipole  frequency is tuned to a single Mie resonance mode inside the nanodiamond. The general case is the simultaneous excitation of several modes. An example is shown for the sake of illustration in Fig. \ref{Fig_Mie_detuned} for the case of a single dipole orientation with a frequency that does not match exactly one resonance. The Purcell factor is then significantly lower and the angular emission pattern becomes more directional.  
In practice, further averaging effects are due to the existence of two orthogonal transition dipoles in the NV center electronic structure \cite{Ref_NV_Deuxdipoles} and  to its broad luminescence spectrum.  Here, we have highlighted the resonant properties of the dielectric particle. This introduces the issue of the interplay between a microcavity and a broad emitter that has been discussed in the context of semiconductor quantum dots and large quality factors  \cite{Auffeves} and in the case of NV center coupled to a low quality Fabry-Perot microcavity \cite{Dumeige}. 
The emission of the NV center in a nanodiamond supporting Mie resonances corresponds  to a broad emitter coupled to a multimode cavity with its free spectral range determined by the radius of the nanoparticle. A further discussion of the emission spectrum modification is beyond the scope of this letter. 

In practice, nanoparticles are usually deposited on the surface of a dielectric substrate which modifies the environment of the particle. We have used a finite element technique in order to  account  for the presence of the dielectric interface in the evaluation of $F^\prime_{\rm p}$. 
Following the experiment reported in Ref. \cite{Beveratos2001}, we  consider a particle of $45\,{\rm nm}$  radius  deposited on  a silica microscope coverslip with $n=1.5$. As the nanoparticles are embedded  in a thin polymer film deposited on the substrate, the particle is considered in a dielectric environment of same refractive index. Such parameters then correspond to an intermediate regime between the electrostatic case and the Mie regime (see Fig. \ref{FigPurcell}). Without accounting for the interface,  the Purcell factor $F^\prime_{\rm p}$ is around $0.1$. The numerical simulation with an interface yields larger values of this parameter, with a maximum $\simeq 0.39 $ for a dipole parallel to the interface. 
Given the uncertainty on particle shape and size  and the fact that the model does not account for any nonradiative decay \cite{Smith2010}, this result can be considered to be in fair agreement with the  approximate experimental value   $\simeq 0.47$   \cite{Beveratos2001}. 

\begin{figure}[Hhbt]

  \epsfig{file=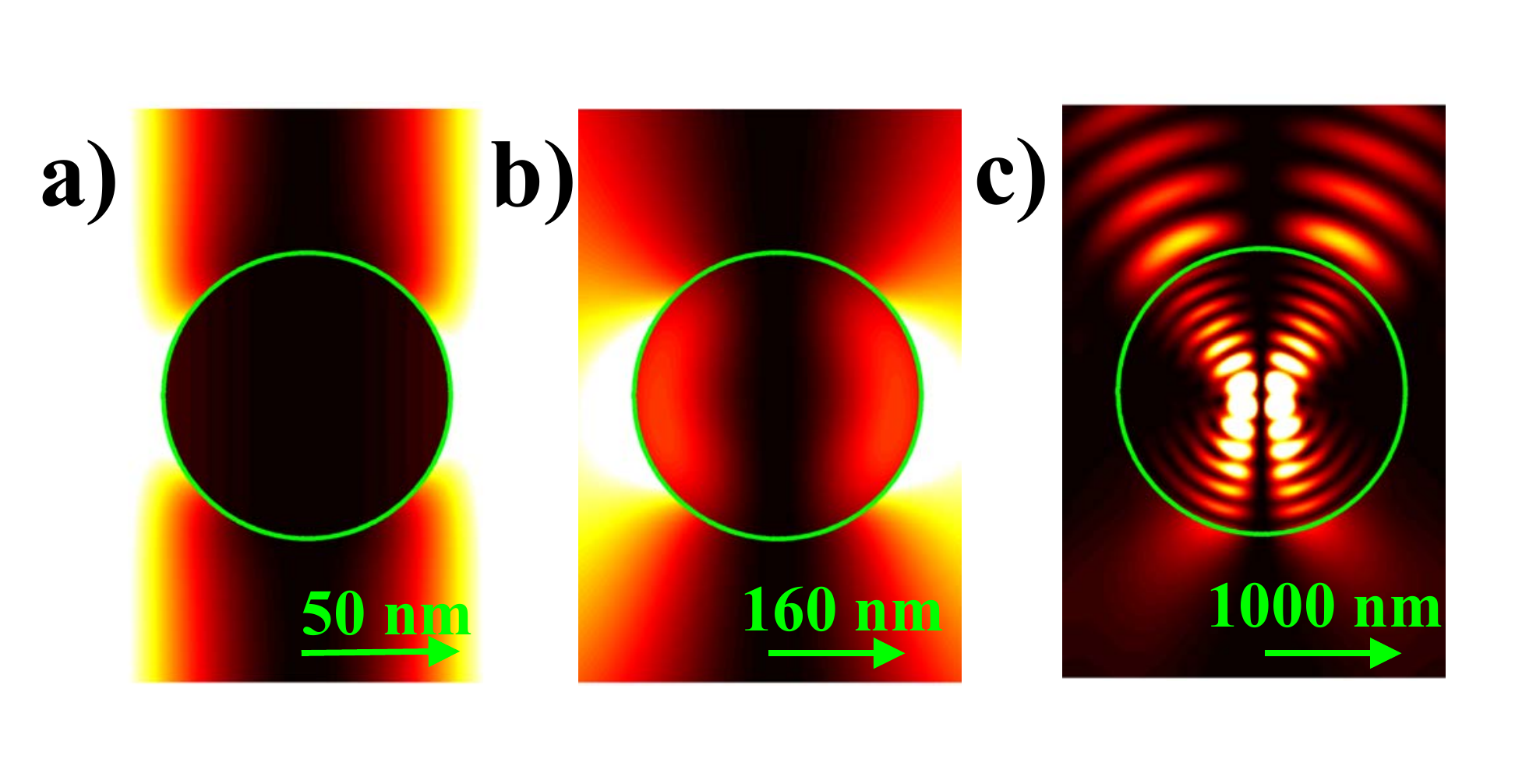, width=0.35\textwidth}
  \caption{Square modulus of the electric field in a dielectric nanosphere illuminated by a structured field, with radii $a= 50\, {\rm nm}$,  $a= 160\, {\rm nm}$ (b) and   $a= 1000\, {\rm nm}$ (c).}
  \label{Fig_STED_nanodiamond}
\end{figure}

We finally address the question of imaging a NV center in the nanodiamond using the STED technique. Since a $6\,{\rm nm}$ record resolution \cite{Rittweger2009a} in a bulk diamond sample has been demonstrated, the technique is expected to be able to image a NV center in a nanoparticle. However, all attempts have failed so far, the  signal appearing as  delocalized over the particle. To explain these results, we again use the view that a dielectric nanoparticle behaves as a nanoantenna. Hence, the structured exciting field used in the STED technique excites first the particle modes which in turn, excite the defect. For a particle much smaller than the wavelength, only the electric dipolar mode is significant. Since this mode is uniform in the particle  (see Fig.\ref{Intensity_nanosphere}(a)) the NV center is equally excited independently of its location. To illustrate this behavior, we display the field produced by a structured beam which illuminates the particle. Following Ref.  \cite{Rittweger2009a}, this beam is generated by applying a phase shift \cite{note}
 to the Fourier components of an input circularly polarized beam which is focused on the particle with a high numerical aperture microscope objective. Figure \ref{Fig_STED_nanodiamond} shows a simulation carried out for NA$=0.9$ where it is seen that  the phase transformation leads to a beam which is depleted in its center. Whereas it is clearly seen that the depleted structure of the incident field penetrates in the sphere of radius $a=1000$ nm, the field becomes uniform in the particle of sub-wavelength size. This can be understood from the previous discussion. Indeed the NV center excitation is mediated by the mode of the particle which is the dipolar one for small particles. Since the electric field of the dipolar mode is uniform in the particle, the spatial structure of the incident field is lost so that the STED resolution is limited by the particle size.

In summary, we have discussed how the emission properties of a NV center in a diamond nanoparticle can be analysed using the concept of nanoantenna. The coupling of a NV center to electromagnetic fields is mediated by the modes of the particle in the Mie regime and  is dominated by  dielectric screening in the case of a sub-wavelength particle. This simple mechanism provides a unified picture of the NV optical properties as a function of the nanodiamond size. More generally, the concept of dielectric nanoantenna can be applied to any dielectric nanostructure containing a luminescent center.
 
This work was supported by Triangle de la Physique contract 2008-057T. We are grateful to Vincent Jacques for many helpful discussions. 

%
%

\kern -6 mm


\begin{thebibliography}{99}
\bibitem{note} A Fourier component $\mathbf{E}(k_x,k_y)$ of an incident Gaussian beam is multiplied by  $\exp[i\phi]$ with $\tan(\phi)=k_x/k_y$.
\end{thebibliography}
\end{document}